\documentclass[aps,twocolumn,superscriptaddress,showpacs]{revtex4}
\usepackage[dvips]{graphicx}
\newcommand{\nc}{\newcommand}		
\newcommand{\renc}{\renewcommand}	
\nc{\vc}[1]	{\mbox{\boldmath $#1$}}	
\nc{\del}       {\partial}              
\nc{\bra}       {\langle}               
\nc{\ket}       {\rangle}               
\nc{\bras}[1]   {\langle #1|}           
\nc{\kets}[1]   {|#1\rangle}            
\nc{\mapleft}[1]{			
 \smash{\mathop{\,			%
  \hbox to 1.5cm{\rightarrowfill}\, }\limits_{#1}}}

\def\JL#1#2#3#4{ {{\rm #1}}\ \textbf{#2}, #4 (#3)}  
\nc{\PR}[3]     {\JL{Phys. Rev.}{#1}{#2}{#3}}
\nc{\PRC}[3]    {\JL{Phys. Rev.~C}{#1}{#2}{#3}}
\nc{\PRA}[3]    {\JL{Phys. Rev.~A}{#1}{#2}{#3}}
\nc{\PRL}[3]    {\JL{Phys. Rev. Lett.}{#1}{#2}{#3}}
\nc{\NP}[3]     {\JL{Nucl. Phys.}{#1}{#2}{#3}}
\nc{\NPA}[3]    {\JL{Nucl. Phys.}{A#1}{#2}{#3}}
\nc{\PL}[3]     {\JL{Phys. Lett.}{#1}{#2}{#3}}
\nc{\PLB}[3]    {\JL{Phys. Lett.~B}{#1}{#2}{#3}}
\nc{\PTP}[3]    {\JL{Prog. Theor. Phys.}{#1}{#2}{#3}}
\nc{\PTPS}[3]   {\JL{Prog. Theor. Phys. Suppl.}{#1}{#2}{#3}}
\nc{\PRP}[3]   {\JL{Phys. Rep.}{#1}{#2}{#3}}
\nc{\JP}[3]     {\JL{J. of Phys.}{#1}{#2}{#3}}
\nc{\andvol}[3] {{\it ibid.}\JL{}{#1}{#2}{#3}}

\nc{\mydraft}	{\setlength{\topmargin}{-1.5cm}}
\mydraft

\begin{document}
\title{
Systematic study of $^{9,10,11}$Li with the tensor and pairing correlations
}

\author{Takayuki Myo
\footnote{E-mail address: myo@rcnp.osaka-u.ac.jp}
}
\affiliation{
Research Center for Nuclear Physics (RCNP),
Ibaraki, Osaka 567-0047, Japan}

\author{Yuma Kikuchi
\footnote{E-mail address: yuma@nucl.sci.hokudai.ac.jp}
}
\affiliation{
Division of Physics, Graduate School of Science,
Hokkaido University, Sapporo 060-0810, Japan.}

\author{Kiyoshi Kat\=o
\footnote{E-mail address: kato@nucl.sci.hokudai.ac.jp}
}
\affiliation{
Division of Physics, Graduate School of Science,
Hokkaido University, Sapporo 060-0810, Japan.}

\author{Hiroshi Toki
\footnote{E-mail address: toki@rcnp.osaka-u.ac.jp}
}
\affiliation{
Research Center for Nuclear Physics (RCNP),
Ibaraki, Osaka 567-0047, Japan}

\author{Kiyomi Ikeda
\footnote{E-mail address: k-ikeda@postman.riken.go.jp}
}
\affiliation{
RIKEN Nishina Center, 2-1 Hirosawa, Wako, Saitama 351-0198, Japan}

\date{\today}

\begin{abstract}
We make a systematic study of Li isotopes ($A=9,10,11$) in the tensor optimized shell model for $^9$Li and treat the additional valence neutrons in the cluster model approach by taking into account the Pauli-blocking effect caused by the tensor and pairing correlations.
We describe the tensor correlations in $^9$Li fully in the tensor-optimized shell model, where the variation of the size parameters of the single particle orbits is essential for getting strong tensor correlations.
We have shown in our previous study that in $^{10,11}$Li the tensor and pairing correlations in $^9$Li are Pauli-blocked by additional valence neutrons, which make the $p$-shell configurations pushed up in energy.
As a result, the $s^2$ valence neutron component increases to reveal the halo structure of $^{11}$Li and the inversion phenomenon of the single particle spectrum in $^{10}$Li arises.
Following the previous study, we demonstrate the reliability of our framework by performing a detailed systematic analysis of the structures of $^{9,10,11}$Li, 
such as the charge radius, the spatial correlation of halo neutrons of $^{11}$Li and the electromagnetic properties of Li isotopes.
The detailed effects of the Pauli-blocking on the spectroscopic properties of $^{10,11}$Li are also discussed.
It is found that the blocking acts strongly for the $^{11}$Li ground state rather than for $^{10}$Li and for the dipole excited states of $^{11}$Li, which is mainly caused by the interplay between the tensor correlation in $^9$Li and the halo neutrons.
The results obtained in these analyses clearly show that the inert core assumption of $^9$Li is not realistic to explain the anomalous structures observed in $^{10,11}$Li.
For the dipole excitation spectrum of $^{11}$Li, the effect of the final state interactions is discussed in terms of the dipole strength function.
\end{abstract}


\maketitle 

\section{Introduction}\label{intro}

The halo structure in $^{11}$Li was completely outside of the common sense of nuclear physics \cite{Ta85}.  Because of this reason, many experiments have been performed on $^{11}$Li and the surrounding nuclei in order to understand the physics causing this interesting phenomenon.  We summarize here the available experimental data accumulated for $^{11}$Li and surrounding nuclei until now:
a) The halo neutrons have almost equal amount of the $s$-wave component with respect to the $p$-wave component \cite{Si99}.  
b) More precise measurements and analyses of r.m.s. radius are obtained \cite{To97,Do06}.
c) The $E1$ strength distribution has a large peak near the threshold \cite{Ie93,Sh95,Zi97,Na06}.  
d) The charge radius is larger than that of $^9$Li \cite{Sa06,Pu06}. 
e) The electromagnetic properties are observed to reveal the exotic structure of $^{11}$Li\cite{Ar94}.

The challenge for the theoretical work is the large $s$-wave component for the halo neutrons,
which means that the shell gap at $N=8$ has to disappear by some reason.
However, the mean field treatment of a central force is not able to provide the disappearance of the $N=8$ shell gap, because the $N=8$ magic number is produced by the shell structure due to the central potential.
Hence, there have been many theoretical studies performed for $^{11}$Li \cite{To90b,Es92,To94,De97,Ba97,Mu98,Va02,My02,Ga02a,Ga02b,Mi06,Br06}, but conventionally, most of them had to assume that the $1s_{1/2}$ single particle state is brought down by hand to the $0p_{1/2}$ state in order to match with the experimental situation. It is therefore the real challenge for theorists to understand this disappearance of the $N=8$ shell gap, to be called the $s$-$p$ shell gap problem.

The halo structure of $^{11}$Li should be related with the states involving the $1s$ state ($1s$ states) and those involving the $0p$-state ($0p$ states) in $^{10}$Li. Several experiments suggest that non-resonant, virtual $1s$-states are located close to the $^9$Li+$n$ threshold energy together with the nearby but higher $0p$ states \cite{Th99,Cha01,Je06}.
The formation of the low-lying virtual $1s$-state in $^{10}$Li is definitely important to understand the inversion phenomenon of the $1s$ and $0p$ states in the $N=7$ isotones. However, this inversion phenomenon is still an unsolved problem \cite{Cha01} in addition to the halo structures of $^{11}$Li.  
Hence, we need an important mechanism to explain the inversion phenomenon of the $s$-$p$ states and the virtual $1s$-state in $^{10}$Li
consistently with the structures of $^{11}$Li.

It is well known by now that the tensor interaction plays an important role for light nuclei.  The deuteron is bound due to the strong coupling between the $s$-state and the $d$-state due to the tensor interaction.  In $^4$He, the tensor interaction produces a strong coupling of the $s$-state and the $d$-state in the relative wave function and eventually the contribution of the tensor interaction becomes energetically comparable to or even more than that of the central interaction \cite{Ak86,Ka01}.  Very recently, the Argonne-Illinois group made a thorough study for light nuclei until mass number of about A=10 using the Green's function Monte Carlo method in the few-body framework.  They take the bare $NN$ interaction including the tensor interaction with a small but important three body interaction for quantitative reproduction of light nuclei \cite{Pi04}.  They found not only very good reproduction of binding energies, but also an important fact that the pion exchange interaction (predominantly tensor interaction) provides about 70$\sim$80\% of the two-body attraction for those light nuclei. Hence, a proper treatment of the tensor interaction seems essential for the understanding of the nuclear structure.

In our recent study \cite{My05,My07}, we have developed a theoretical framework of the tensor-optimized shell model (TOSM) to treat the tensor interaction in the shell model basis explicitly.  This finding is very important since we now have a tool to systematically understand the role of the tensor interaction and its correlation in the shell model language.  The first place to use this method was the $^4$He+$n$ system, where we are able to show that the tensor correlation provides a considerable amount of the $p_{1/2}$-$p_{3/2}$ splitting energy in $^5$He \cite{My05,Ik06,Na59,Te60}.  Hence, we can apply this method for the present shell gap problem.
In our previous paper\cite{My07b}, we presented our basic idea and the formulation to treat the tensor correlation in the $^9$Li+$n$+$n$ system. 
In the present paper, we shall demonstrate the reliability and the validity of this framework for Li isotopes by showing various physical quantities observed not only for $^{11}$Li, but also for the neighboring isotopes $^9$Li and $^{10}$Li.
We have applied TOSM to the $^9$Li core to describe the tensor correlation explicitly.
It is found that the $(0s_{1/2})^{-2}(0p_{1/2})^2$ of the $2p$-$2h$ excitation of a proton-neutron pair has a special importance in describing the 
relative $sd$ coupling of the tensor interaction and the tensor correlation in $^9$Li, which is the same result as obtained for $^4$He \cite{My07,Su04,Og06,Og07}. 
For $^{11}$Li, we solved the coupled $^9$Li+$n$+$n$ three-body problem with the tensor correlation in $^9$Li\cite{My07b}.
The tensor correlation arising from the $2p$-$2h$ excitations in $^9$Li is Pauli-blocked by additional two neutrons, which makes the $p$-shell configurations pushed up in energy. 
As a result, the $s$-$p$ shell gap becomes prominently small and the $s^2$ component for the last neutrons increases largely around 25\%.

The tensor correlation is, however, not enough for quantitative description of the $s$-$p$ shell gap problem.  We have to add further the paring correlation among the $p$-shell states.  Hence, we have solved a coupled problem with the tensor and pairing correlations in $^9$Li. 
We confirmed then that the neutron pairing correlation in $^9$Li works coherently with the tensor correlation and produces the similar blocking effect in $^{11}$Li as the case of the tensor correlation \cite{Sa93,Ka99,My02}.  
As the final result with the tensor and pairing correlations, the $s$-$p$ shell gap disappears and the $s^2$ component for last neutrons naturally increases as around 50\% which is enough to produce the halo structure of $^{11}$Li.  Our theory further reproduces the large charge radius and the Coulomb breakup strength of $^{11}$Li, and the inversion phenomena of $s$-$p$ states in $^{10}$Li.

So far, there have been appearing a number of experimental data on $^{11}$Li and neighboring nuclei.  Hence, in this paper, we shall further perform a detailed  and systematic analysis of the structures of the Li isotopes on the basis of the tensor optimized shell model \cite{My07b}.
We investigate the Pauli-blocking effect on various physical quantities of $^{10,11}$Li, such as 
the $^{10}$Li spectra of $s$ and $p$ states and the ground state properties of $^{11}$Li including the spatial correlation of halo neutrons. The charge radius, the electromagnetic properties of $^9$Li and $^{11}$Li are also investigated.
For the three-body Coulomb breakup of $^{11}$Li, we confirm the low-energy peak in the strength and the origin of this peak is investigated from the view point of the final state interactions.
The difference of the Pauli-blocking effects in $^{10}$Li and $^{11}$Li is also discussed, 
which is caused by the coupling of the $^9$Li core excitation and the halo neutrons.

This paper is arranged as follows.  In Sec.~\ref{model}, we explain the wave functions of $^{9,10,11}$Li, which describe the tensor and pairing correlations explicitly.
In Sec.~\ref{result}, based on the obtained results of $^9$Li, we show the role of the Pauli-blocking effects coming from the tensor and pairing correlations on $^{10,11}$Li and show various properties of these nuclei.
A Summary is given in Sec.~\ref{summary}.

\section{Model}\label{model}

\subsection{The tensor optimized shell model for $^{9}$Li}

We first explain the tensor optimized shell model (TOSM) for $^9$Li \cite{My07b}.
The Hamiltonian for $^9$Li is given as,
\begin{eqnarray}
    H(\mbox{$^9$Li})
&=& \sum_{i=1}^9{t_i} - t_G  + \sum_{i<j} v_{ij}\ .
    \label{H9}
\end{eqnarray}
Here, $t_i$, $t_G$, and $v_{ij}$ are the kinetic energy of each nucleon, the center-of-mass (c.m.) term and the two-body $NN$ interaction consisting of central, spin-orbit, tensor and Coulomb terms, respectively. The wave function of the $3/2^-$ state in $^9$Li is described in the tensor-optimized shell model \cite{My05,My07}. We express $^9$Li by a multi-configuration,
\begin{eqnarray}
  \Psi(^{9}\mbox{Li})=\sum_i^N a_i\, \Phi^{3/2^-}_i,
  \label{WF9}
\end{eqnarray}
where we take the $0p$-$0h$ state of $^9$Li as $(0s)^4(0p_{3/2})^5$ and consider up to the $2p$-$2h$ excitations from the $^9$Li core state configuration within the $0p$ shell for $\Phi^{3/2^-}_i$ in a shell model type wave function, and $N$ is the configuration number. Based on the previous study \cite{My05,My07,My07b}, we adopt the spatially modified Gaussian function as a single-particle orbit and treat the length parameter $b_\alpha$ of each orbit $\alpha$ ($0s$, $0p_{1/2}$ and $0p_{3/2}$) as variational parameters. 

In the tensor-optimized shell model, we solve the following variational equations with respect to the total wave function $\Psi(^9{\rm Li})$ as,
\begin{eqnarray}
    \frac{\del \bra\Psi| H(\mbox{$^9$Li}) - E |\Psi \ket} {\del a_{i}}
=   0,~~\quad
    \frac{\del \bra\Psi| H(\mbox{$^9$Li}) - E |\Psi \ket} {\del b_{\alpha}}
=   0.~
    \label{TOSM}
\end{eqnarray}
Here, $E$ is the total energy of $^9$Li.
We determine $\{a_i\}$ and the length parameters $\{b_{\alpha}\}$ of three orbits. 
We can optimize the radial form of single-particle orbits appropriately in the wave function in Eq.~(\ref{WF9}), such as to describe the spatial shrinkage of the particle state, which is important in realizing the tensor correlation \cite{My05,My07,My07b,To02,Su04,Og06,Og07}.
We add further the neutron paring correlation for the $p$-shell orbits.

In the description of the tensor correlation, in principle, we can work out a large space to include the full space effect of the tensor force by taking $2p$-$2h$ states with very high angular momenta \cite{My07}. In order to avoid large computational efforts without loss of the physical importance in the result, we restrict the $2p$-$2h$ shell model states within the $p$-wave states for the description of $^9$Li with the single Gaussian basis.  Instead, we increase the tensor matrix elements by 50\%.  
The justification of this enhancement factor is explained as follows, which has also been discussed in Ref.\cite{My07}.
In TOSM, 
addition of the higher configurations beyond the $p$-shell increases the total amount of the $2p$-$2h$ components in the wave function, 
while these configurations share the total $2p$-2$h$ components and reduce the $2p$-$2h$ components including the $p$-shell configurations\cite{My07}.
On the other hand, the spatial behaviour of each single particle basis can be improved also using the Gaussian expansion method (GEM), 
from the compact Gaussian form to more suitable one. This GEM effect was found to enhance each $2p$-$2h$ component\cite{My07,My07b}.
Both effects can be included in the wave function of the core in TOSM.
When the Pauli-blocking effect is considered by adding an extra neutron into the particular occupied orbit of neutron in the core,
the mixing of higher configurations reduces this blocking because of the sharing of the $2p$-$2h$ components.  
On the other hand, the GEM effect enhances the blocking, 
because the Pauli-blocking is proportional to the spatial overlap of the neutron wave functions between the core and the extra neutron. 
Actually, it was checked that in the scattering problem of the $^4$He+$n$ system, the combined effect of the higher configurations 
and GEM on the Pauli-blocking is simulated using the enhanced tensor matrix elements with a single Gaussian basis\cite{Ik06,My07b}.
Therefore, in the present study, we adopt this approximation of the tensor correlation for $^9$Li.

We discuss here the interaction for $^9$Li; $v_{ij}$ in $H(^9{\rm Li})$, where we use the shell model wave functions for the $^9$Li core in Eq.~(\ref{WF9}). Since our main interest in this work is to investigate the role of the tensor interaction on the two-neutron halo formation, we describe the tensor correlation in addition to the pairing correlation in $^9$Li. We use the GA interaction proposed by Akaishi \cite{My05,Ak04,Ik04} for $v_{ij}$ in Eqs.~(\ref{H9}), (\ref{H11}) and (\ref{H10}). 
This effective interaction GA has the tensor force obtained from the $G$-matrix calculation using the AV8$^\prime$ bare force 
keeping a large momentum space \cite{Ak04,Ik04}. 
Actually, it was shown in the previous paper\cite{My07} that the tensor force matrix element using GA for the relative $sd$ coupling
almost reproduces the results of AV8$^\prime$. 
This means that we employ the effective tensor force in GA, which almost has the same strength of the bare tensor force.
In GA, the obtained $^9$Li wave function in Eq.~(\ref{WF9}) shows smaller matter radius than the observed one due to the high momentum component produced by the tensor correlation \cite{My05,Su04,Og06}.
Hence, we phenomenologically adjust the central force, which is done by changing the second range of the central force by reducing the strength by $21.5\%$ and increasing the range by 0.185 fm to reproduce the observed binding energy (45.3 MeV) and the matter radius of $^9$Li\cite{My05,My07,My07b}.

\subsection{$^{10}$Li as $^9$Li+$n$ system in TOSM}

For $^{10}$Li, the Hamiltonian for $^9$Li+$n$ is given as
\begin{eqnarray}
  H(\mbox{$^{10}$Li})
&=&H(\mbox{$^9$Li})
+   T_0 + T_1 - T^{(2)}_G + V_{cn},
    \label{H10}
\end{eqnarray}
where $H(\mbox{$^9$Li})$, $T_i$ and $T^{(2)}_G$  are the internal Hamiltonian of $^9$Li given by Eq.~(\ref{H9}), the kinetic energies of each cluster ($T_0$ for $^9$Li, the same as $t_G$ in Eq.~(\ref{H9})), the c.m. term of two cluster systems, respectively. ${V}_{cn}$ is the $^9$Li core-$n$ interaction. The wave function of $^{10}$Li with the spin $J$ is given as
\begin{eqnarray}
    \Psi^{J}(^{10}{\rm Li})
&=& \sum_i^N{\cal A}\left\{ [\Phi^{3/2^-}_i, \chi^{J_0}_i(n)]^{J} \right\}.
    \label{WF10}
\end{eqnarray}
We use the $^9$Li wave functions in TOSM and obtain coupled differential equations for the neutron wave function, $\chi^{J_0}(n)$
in order to describe the relative motion between $^9$Li and neutron, where $J_0$ is the total angular momentum of the last neutron part of $^{10}$Li. To obtain the total wave function, $\Psi^{J}(^{10}{\rm Li})$, we use the orthogonality condition model (OCM) \cite{To90b,Ka99,Sa69} to treat the antisymmetrization between the last neutron and neutrons in $^9$Li. In OCM, the neutron wave functions $\chi$ are imposed to be orthogonal to the occupied orbits by neutrons in $^9$Li, which depends on the configuration $\Phi^{3/2^-}_i$ in Eq.~(\ref{WF9}). We obtain the following coupled Schr\"odinger equations with OCM for the set of the wave function, $\{\chi_i^{J_0}(n)\}$, where $i=1,\cdots,N$:
\begin{eqnarray}
\lefteqn{\hspace*{-2.8cm}
\sum_{j=1}^N \left[ \left(T_{\rm rel}^{(2)} + V_{cn} + \Lambda_i \right) \delta_{ij} + h_{ij}(^9{\rm Li})\right]}
\nonumber
\\
\times~\chi_j^{J_0}(n)&=&E\ \chi_i^{J_0}(n),
\label{OCM10}
\\
\Lambda_i
&=& \lambda \sum_{\alpha\in \Phi_i(^{9}{\rm Li})} |\psi_\alpha \ket \bra \psi_\alpha|,
\label{LAM10}
\end{eqnarray}
where $h_{ij}(^9{\rm Li})=\bra \Phi_i^{3/2^-} | H(^{9}{\rm Li}) | \Phi_j^{3/2^-} \ket$. $T^{(2)}_{\rm rel}$ is the kinetic energy of the relative motion for $^{10}$Li. $\Lambda_i$ is the projection operator to remove the Pauli forbidden states $\psi_\alpha$ from the relative wave functions of last neutron\cite{Ka99,Ku86}, where $\psi_\alpha$ is the occupied single particle wave function of the orbit $\alpha$ in $^9$Li. This $\Lambda_i$ depends on the configuration $\Phi^{3/2^-}_i$ of $^{9}$Li
and plays an essential role to produce the Pauli-blocking in $^{10}$Li. 
The value of $\lambda$ is taken as large as $10^6$~MeV in the present calculation in order to project out the components of the Pauli forbidden states into an unphysical energy region.
Here, we keep the length parameters {$b_\alpha$} of the single particle wave functions as those obtained for $^9$Li. 
The detailed treatment of the orthogonality condition was explained in Ref.~\cite{My07b}.

The $^9$Li-$n$ potential, $V_{c n}$, in Eq. (\ref{H10}) is given by folding an effective interaction, the MHN $G$-matrix interaction\cite{Ha71,Fu80,To90b,Ao06,My07b}. Any state-dependence is not introduced in the $^9$Li-$n$ potential. We introduce one parameter, $\delta$, which is the second-range strength of the MHN interaction to describe the starting energy dependence dominantly coming from the tensor interaction in the $G$-matrix calculation \cite{Ao06,Fu80,Ya74}. In the present calculation, we chose this $\delta$ parameter $0.1745$ to reproduce the two-neutron separation energy $S_{2n}$ of $^{11}$Li as 0.31 MeV, since experimental situation is not so clear for the $^{10}$Li spectra due to the unbound system.

\subsection{$^{11}$Li as $^9$Li+$n$+$n$ system in TOSM}

For $^{11}$Li, the Hamiltonian of the $^9$Li+$n$+$n$ three-cluster system, is given as
\begin{eqnarray}
  H(\mbox{$^{11}$Li})
&=&H(\mbox{$^9$Li})
+   \sum_{k=0}^2{T_k} - T^{(3)}_G
+   \sum_{k=1}^2V_{cn,k}
\nonumber
\\
&&+~V_{nn}, 
    \label{H11}
\end{eqnarray}
where $T^{(3)}_G$ is the kinetic energy of the c. m. term of three cluster systems. ${V}_{cn,k}$ are the $^9$Li core-$n$ interaction ($k=1,2$) and $V_{nn}$ is the interaction between the last two neutrons. The wave functions of $^{11}$Li with the spin $J^\prime$, are given as
\begin{eqnarray}
    \Psi^{J^\prime}(^{11}{\rm Li})
&=& \sum_i^N{\cal A}\left\{ [\Phi^{3/2^-}_i, \chi^{J^\prime_0}_i(nn)]^{J^\prime} \right\}.
    \label{WF11}
\end{eqnarray}
We obtain coupled differential equations for the neutron wave functions $\chi^{J^\prime_0}(nn)$, where $J^\prime_0$ is the total angular momenta of the last neutrons. To obtain the total wave function $\Psi^{J^\prime}(^{11}{\rm Li})$, we use the orthogonality condition model (OCM)\cite{To90b,My02,Ao06} to treat the antisymmetrization between last neutrons and neutrons in $^9$Li. In OCM, the neutron wave functions $\chi$ are imposed to be orthogonal to the occupied orbits by neutrons in $^9$Li, which depends on the configuration $\Phi^{3/2^-}_i$ in Eq.~(\ref{WF9}). We obtain the following coupled Schr\"odinger equations with OCM for the set of the wave functions $\{\chi_i^{J^\prime_0}(nn)\}$ for $^{11}$Li, where $i=1,\cdots,N$:
\begin{eqnarray}
\lefteqn{\hspace*{-3.7cm}
\sum_{j=1}^N \left[ \left(T_{\rm rel}^{(3)} + \sum_{k=1}^2(V_{cn,k}+\Lambda_{i,k})+V_{nn} \right) \delta_{ij} + h_{ij}(^9{\rm Li})\right]
}
\nonumber
\\
\times~\chi_j^{J^\prime_0}(nn)&=&E\ \chi_i^{J^\prime_0}(nn),
\nonumber
\\
\label{OCM11}
\end{eqnarray}
where $h_{ij}(^9{\rm Li})=\bra \Phi_i^{3/2^-} | H(^{9}{\rm Li}) | \Phi_j^{3/2^-} \ket$. $T^{(3)}_{\rm rel}$ is the total kinetic energies consisting of the relative motions for $^{11}$Li. 
$\Lambda_{i,k}$ is the projection operator for each last neutron with an index $k$.

We describe the two neutron wave functions $\chi(nn)$ in Eq.~(\ref{OCM11}) precisely adopting a few-body approach of the hybrid-TV model\cite{To90b,My02,Ao95}:
\begin{eqnarray}
      \chi^{J^\prime_0}_i(nn)
&=&   \chi^{J^\prime_0}_i(nn,\vc{\xi}_V)+\chi^{J^\prime_0}_{i}(nn,\vc{\xi}_T),
      \label{TV}
\end{eqnarray}
where $\vc{\xi}_V$ and $\vc{\xi}_T$ are V-type and T-type coordinate sets of a three-body system, respectively. The radial part of the relative wave function is expanded with a finite number of Gaussian basis functions centered at the origin, which can describe the loosely binding wave function of neutron halos\cite{Ao06}.

Here, we discuss the coupling between $^9$Li and the last neutrons, whose details were already explained in the previous paper\cite{My02,My07b}. We consider the $^{11}$Li case. Asymptotically, when the last two neutrons are independently far away from $^9${Li} ($\vc{\xi}_{V,T}\to\infty$), any coupling between $^9$Li and two neutrons disappears and $^9$Li becomes its isolated ground state. Namely, the mixing coefficients $\{a_i\}$ are the same as those obtained in Eq.~(\ref{WF9}):
\begin{eqnarray}
	\Phi^{J^\prime}(^{11}{\rm Li})
\mapleft{\vc{\xi}_{V,T}\to\infty}
	\left[ \Psi(^{9}\mbox{Li}), \chi^{J^\prime_0}(nn) \right]^{J^\prime},
	\label{asympt}
	\\
 \Psi(^{9}\mbox{Li})=\sum_i^N a_i\, \Phi^{3/2^-}_i,
\end{eqnarray}
where $\{\chi^{J^\prime_0}_i(nn)\}$ in Eq.~(\ref{WF11}) satisfies the following asymptotic forms for $i=1,\cdots,N$ 
\begin{eqnarray}
    \chi^{J^\prime_0}_i(nn)
&\mapleft{\vc{\xi}_{V,T}\to\infty} & a_i \cdot \chi^{J^\prime_0}(nn).
    \label{asympt2}
\end{eqnarray}
This equation implies that the asymptotic wave function of two neutrons $\chi^{J^\prime_0}_i(nn)$ is decomposed into the internal amplitude $a_i$ of $^9$Li and the relative wave function $\chi^{J^\prime_0}(nn)$. Equations~(\ref{asympt})-(\ref{asympt2}) give the boundary condition of the present coupled three-body model of $^{11}$Li. Contrastingly, when the two neutrons are close to $^9$Li, two neutrons dynamically couple to the configuration $\Phi^{3/2^-}_i$ of $^9$Li satisfying the Pauli-principle. 
This coupling depends on the relative distance between $^9${Li} and the neutrons and changes $\{a_i\}$ of $^9$Li from those of the $^9${Li} ground state.
The structure change of $^9$Li inside $^{11}$Li is determined variationally to minimize the energy of the $^{11}$Li ground state.
The dynamical effect of the coupling arising from the Pauli-blocking is explained in the results in detail.

For the potential $V_{nn}$ between the last two neutrons, we take a realistic interaction AV8$^\prime$ in Eq.~(\ref{H11}). Our interest is to see the $n$-$n$ correlation in the two-neutron halo structure. Hence we solve two-neutron wave function $\chi(nn)$ with no assumption in the hybrid-TV model in Eq.~(\ref{TV}), which can describe the short range correlation under the realistic interaction AV8$^\prime$.

\section{Results}\label{result}

\subsection{$^9$Li}

\begin{figure}[t]
\centering
\includegraphics[width=8.1cm,clip]{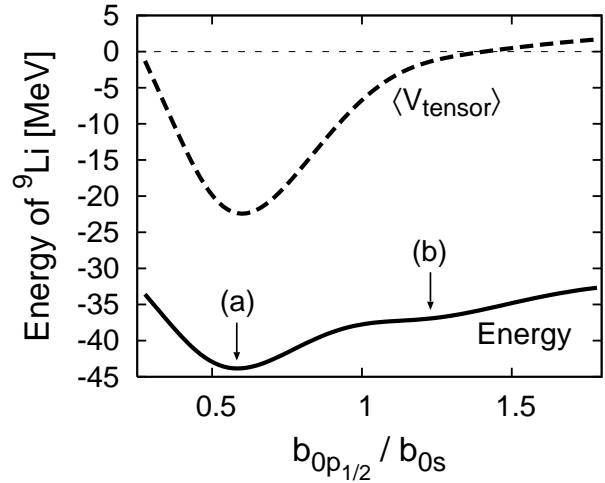}
\caption{Energy surface of $^9$Li (solid line) with respect to the length parameters of $0p_{1/2}$ orbit normalized by that of $0s$ orbit.  The two minima indicated by (a) and (b) correspond to the states due to the tensor correlation and the paring correlation, respectively.
The dashed line is the tensor force contribution.}
\label{9Li_ene}
\end{figure}

We show first the results of $^9$Li in TOSM, which is the basis for heavier Li isotopes.
As the first step, we search for the energy minima of $^9$Li with respect to the length parameters $\{b_\alpha\}$ by solving Eq.~(\ref{TOSM}). 
Fig. \ref{9Li_ene} shows the result of the energy surface with respect to the $b_{0p_{1/2}}$, 
where $b_{0s}$ and $b_{0p_{3/2}}$ are already optimized as 1.45 fm and 1.8 fm, respectively. 
We found two energy minima, (a) having a small value (0.85 fm) of $b_{0p_{1/2}}$
and (b), which is a rather shallow local minimum, having a large value of $b_{0p_{1/2}}$ as 1.8 fm,
the same length of $b_{0p_{3/2}}$. 
The contribution of the tensor force $\langle V_{\rm tensor}\rangle$ is also shown.
As was explained in the previous paper\cite{My07b}, 
the properties of each minimum are characterized with their dominant $2p$-$2h$ configurations.
At the lowest minimum point (a), the dominant $2p$-$2h$ configuration is given by $(0s)^{-2}_{10}(0p_{1/2})^2_{10}$,
where the subscripts 01 or 10 represent spin and isospin for the two-nucleon pair, respectively.
This configuration provides the largest tensor force contribution with a spatial shrinkage of the $0p_{1/2}$ orbit
and produces the high momentum component, similar to the results in Refs.~\cite{My05,My07}. 
At this energy minimum, $\langle V_{\rm tensor} \rangle$ is about three times as large as 
in the ordinary shell model case with $b_{0p_{1/2}}/b_{0s}$=1.
At the second minimum point (b), $(0p_{3/2})^{-2}_{01}(0p_{1/2})^{2}_{01}$, the $0p$ shell pairing correlation is dominant.
The results indicate that (a) and (b) represent the tensor correlation and the pairing one, respectively.

As the second step, we superpose the above two wave functions at (a) and (b) in order to obtain the $^9$Li wave function 
including the tensor and pairing correlations, simultaneously.
In the superposed $^9$Li wave function, the dominant configurations are as follows with their probabilities 
$0p$-$0h$ with $82.9$ \%, $(0p_{3/2})^{-2}_{01}(0p_{1/2})^2_{01}$ with $9.0$ \% and 
$(0s)^{-2}_{10}(0p_{1/2})^2_{10}$ and $7.2$ \%, respectively. 
From the two kinds of the preferred $2p$-2$h$ configurations, 
the obtained $^9$Li wave function is found to possess both the tensor and pairing correlations.
We show the contributions of the Hamiltonian component in Eq.~(\ref{H9}) in Table \ref{ene9}
and the tensor force gives $-20.7$ MeV.  The contribution from the pairing correlation is $-3.0$ MeV.
We also show the results of the various r.m.s. radius $R$ of matter($m$), proton($p$), neutron($n$), charge($ch$)
for $^9$Li in Table \ref{tab:9}. 
The recent observations of the charge radius is reproduced well in our wave function,
where the radius of proton and neutron is considered.  We present other properties for $^9$Li in comparison with other Li isotopes later.

\begin{table}[t]
\caption{Contributions of the Hamiltonian to the $^9$Li ground state in units of MeV.}
\label{ene9}
\begin{center}
\begin{ruledtabular}
\renc{\baselinestretch}{1.15}
\begin{tabular}{cccccc}
Energy    &  Kinetic  &  Central   & Tensor  & LS      & Coulomb  \\
\noalign{\hrule height 0.5pt}
$-45.3$   &  136.0    &  $-158.8$  & $-20.7$ & $-3.6$  &  1.8     \\
\end{tabular}
\end{ruledtabular}
\end{center}
\end{table}

\begin{table}[t]
\caption{Various r.m.s. radius of $^9$Li given in unit of fm in TOSM and 
those of experiment.}
\label{tab:9}
\begin{center}
\begin{ruledtabular}
\renc{\baselinestretch}{1.15}
\begin{tabular}{c|cc}
                             &  TOSM   &   Experiment                \\
\noalign{\hrule height 0.5pt}
 $R_m$                       & $2.31$  &   2.32$\pm$0.02\cite{Ta88b} \\
                             &         &   2.44$\pm$0.06\cite{Do06}  \\
 $R_{p}$                     & $2.10$  &   2.18$\pm$0.02\cite{Ta88b} \\
 $R_{n}$                     & $2.41$  &   2.39$\pm$0.02\cite{Ta88b} \\
 $R_{ch}$                    & $2.23$  &   2.217$\pm$0.035\cite{Sa06}\\ 
                             &         &   2.185$\pm$0.033\cite{Pu06}\\
\end{tabular}
\end{ruledtabular}
\end{center}
\end{table}

\subsection{Pauli-blocking effects in $^{10}$Li and $^{11}$Li}\label{sec:PB}

\begin{figure}[t]
\centering
\includegraphics[width=8.4cm,clip]{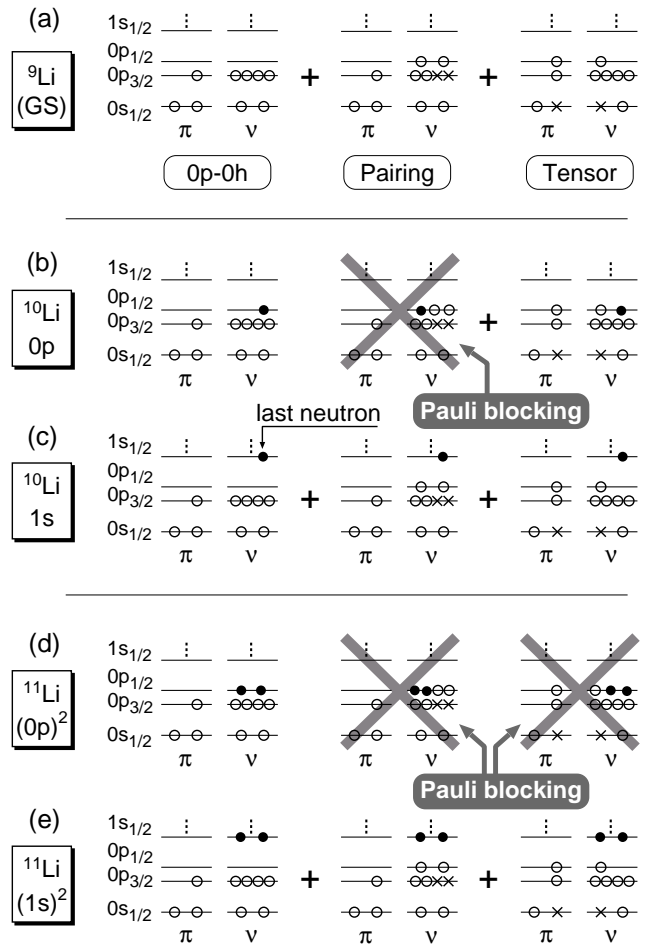}
\caption{Schematic illustration for the Pauli-blocking in $^{10}$Li and $^{11}$Li.
Details are described in the text.}
\label{fig:Pauli}
\end{figure}

Considering the properties of the configuration mixing of $^9$Li,
we discuss the Pauli-blocking effects in $^{10}$Li and $^{11}$Li and their difference as shown in Fig.~\ref{fig:Pauli}.
For (a) in Fig.~\ref{fig:Pauli}, the $^9$Li ground state (GS) is strongly mixed, 
in addition to the $0p$-$0h$ state, with the $2p$-$2h$ states caused by the tensor and pairing correlations.

Let us add one neutron to $^9$Li for $^{10}$Li. 
For (b) in Fig.~\ref{fig:Pauli}, when a last neutron occupies the $0p_{1/2}$-orbit for the $p$-state of $^{10}$Li,
the $2p$-$2h$ excitation of the pairing correlation in $^9$Li are Pauli-blocked.
The tensor correlation is also blocked partially, but not fully by Pauli-principle 
because the $0p_{1/2}$ orbit is not fully occupied by a last neutron. 
Accordingly, the correlation energy of $^9$Li is partially lost inside $^{10}$Li.
Contrastingly, for (c) in Fig.~\ref{fig:Pauli}, the $1s$ state of $^{10}$Li, the Pauli-blocking does not occur and $^9$Li gains its correlation energy fully by the configuration mixing with the $2p$-$2h$ excitations.
Hence, the energy difference between $p$ and $s$ states of $^{10}$Li becomes small to explain the inversion phenomenon\cite{Ka99,My02}.

For $^{11}$Li, let us add two neutrons to $^9$Li.
The similar blocking effect is expected for $^{11}$Li, whose important properties were given in the previous paper\cite{My07b}.
For (d) in Fig.~\ref{fig:Pauli}, when two neutrons occupy the $0p_{1/2}$-orbit,
the $2p$-$2h$ excitations of the tensor and pairing correlations in $^9$Li are Pauli-blocked.
In particular, the blocking of the tensor correlation in $^{11}$Li is expected to work stronger than the $^{10}$Li case,
due to the presence of the last two neutrons in the $p_{1/2}$ orbit.
Accordingly, the correlation energy of $^9$Li is lost inside $^{11}$Li stronger than the $^{10}$Li case.
For (e) in Fig.~\ref{fig:Pauli}, the $(1s)^2$ of two neutrons, the Pauli-blocking does not occur, similar to the $1s$ state of $^{10}$Li.
Hence, the relative energy distance between $(0p)^2$ and $(1s)^2$ configurations of $^{11}$Li becomes small to break the magicity in $^{11}$Li.
It is found that there is a difference in the blocking effects between $^{11}$Li and $^{10}$Li.
It is interesting to examine how this difference affects the $s$-$p$ shell gap problem in these nuclei.
For the pairing correlation, we already pointed out the different blocking effects between $^{10}$Li and $^{11}$Li in the previous study\cite{My02}.
We further consider the blocking effect in the dipole excited states in $^{11}$Li later, which is also different from the ground state case.
In the previous paper\cite{My07b}, we examined that the configuration mixing of the $sd$-shell for $^9$Li would give a small influence 
on the blocking effect on the $(1s)^2$ configuration of $^{11}$Li.

\subsection{$^{10}$Li}

We solve $^{10}$Li in a coupled $^9$Li+$n$ model, in which 
the $^9$Li-$n$ folding potential is determined to reproduce the two-neutron separation energy $S_{2n}$ of $^{11}$Li as 0.31 MeV.
On this condition, we investigate the spectroscopic properties of $^{10}$Li.
In Table \ref{tab:10}, it is shown that using TOSM for $^9$Li, 
the dual $p$-state resonances are obtained near the $^9$Li+$n$ threshold energy.
The dual states come from the coupling of $[(0p_{3/2})_\pi(0p_{1/2})_\nu]_{1^+/2^+}$,
while the experimental uncertainty is still remaining including the spin assignment\cite{Je06}.
The $1^+$ state is predicted at a lower energy than the $2^+$ state due to the attractive effect of the triplet-even $^3E$ channel in the $pn$ interaction.

For the $s$-wave states, their scattering lengths $a_s$ of the $^9$Li+$n$ system show negative values.
In particular, the $2^-$ state shows a large negative value of $a_s$, which is comparable to that of the  $nn$ system 
($-18.5$ fm) with the $^1S_0$ channel, and indicates the existence of the virtual $s$-state near the $^9$Li+$n$ threshold energy.
Therefore the inversion phenomenon in $^{10}$Li is reasonably explained in the present model.
The order of $2^-$ and $1^-$ also comes from the attractive $^3E$ component in the $pn$ interaction.

For comparison, we calculate $^{10}$Li without the core excitations of $^9$Li (``Inert Core''), namely, we adopt only the single $0p$-$0h$ configuration for $^9$Li without the Pauli-blockings explained in Fig.~\ref{fig:Pauli}(d).
In this case, we adjust the $\delta$ parameter to be 0.066.
In Table \ref{tab:10}, the $p$-wave resonances are obtained just above the $^9$Li+$n$ threshold energy, and $a_s$ values show small positive values for both $1^-$ and $2^-$ states, 
which means that the virtual $s$ states are not located near the $^9$Li+$n$ threshold, and the $s$-$p$ shell gap is large.
These results mean that the Pauli-blocking nicely describes the spectroscopic properties of $^{10}$Li.

\begin{table}[t]
\caption{The resonance energies $E_r$ and the decay widths $\Gamma$ of the $p$-wave resonance states (1$^+$ and 2$^+$ states) in $^{10}$Li 
measured from the $^9$Li+$n$ threshold are listed in unit of MeV. 
The scattering lengths $a_s$ of the $s$-states (1$^-$ and 2$^-$ states) are shown in unit of fm.  We show here the two kinds of the results using TOSM and Inert Core for $^9$Li.}
\label{tab:10} 
\begin{center}
\begin{ruledtabular}
\renc{\baselinestretch}{1.15}
\begin{tabular}{c|cc}
                              &    TOSM      &  Inert Core    \\
\noalign{\hrule height 0.5pt}
$(E_r,\Gamma)(1^+)$ [MeV]     & (0.22, 0.09) &  (0.03, 0.005) \\
$(E_r,\Gamma)(2^+)$ [MeV]     & (0.64, 0.45) &  (0.33, 0.20)  \\
$a_s(1^-)$ [fm]               &  $ -5.6$     &    1.4         \\
$a_s(2^-)$ [fm]               &  $-17.4$     &    0.8         \\
\end{tabular}
\end{ruledtabular}
\end{center}
\end{table}

\begin{figure}[t]
\centering
\includegraphics[width=8.0cm,clip]{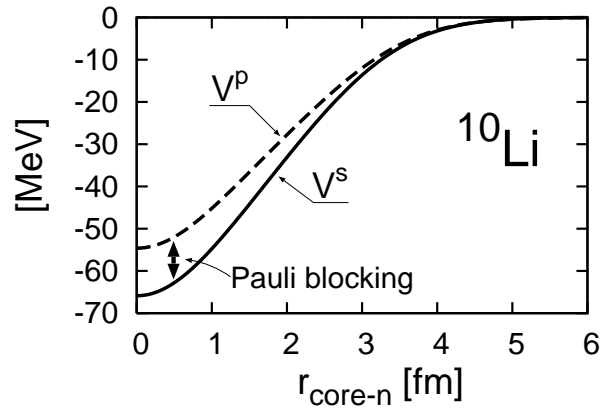}
\caption{Effective $^9$Li-$n$ potentials, $V^s$ and $V^p$ obtained by renormalizing the core excitation in $^{10}$Li,
where the $^9$Li-$n$ potentials are transformed into equivalent ones using the inert $^9$Li core.
$V^s$ is taken to be the same as the original folding potential $V_{cn}$.}
\label{potential10}
\end{figure}

Here, we visualize the Pauli-blocking effect occurred in $^{10}$Li using the effective $^9$Li-$n$ potential. 
In Fig.~\ref{potential10}, we depict the effective $^9$Li-$n$ potentials for the $s$-wave ($V^s$) and for the $p$-wave ($V^p$) of $^{10}$Li, which are derived as follows;
We use the inert description of the $^9${Li} core, namely, no Pauli-blocking from the tensor and pairing correlations
and renormalize the blocking effect into $V^s$ and $V^p$.
For the $s$-wave, $V^s$ is given by the original folding potential $V_{cn}$ in Eq.~(\ref{H10}), because of no blocking from two correlations. 
On the other hand, for the $p$-wave, we reduce the strength of $V_{cn}$ to simulate the blocking effect 
by reproducing the resonance energy of the $1^+$ state (0.22 MeV) of $^{10}$Li as obtained in TOSM shown in Table \ref{tab:10}.
This procedure is similarly done as explained in Refs.~\cite{Ka99,My02}.
In this manner, we obtain the state-dependent interactions with a deeper one for $s$-wave.
The difference, $\Delta_{PB}^n$, between $V^p$ and $V^s$ reflects the strength of the Pauli-blocking (PB):
\begin{eqnarray}
  V^p(r) &\sim& V_{cn}(r) + \Delta_{PB}^n(r),
  \\
  V^s(r) &\sim& V_{cn}(r),
\\
\Delta_{PB}^n(r)&\sim&  V^p(r)-V^s(r).
\end{eqnarray}
$\Delta_{PB}^n$ depends on the relative distance $r$ between the $^9$Li core and a neutron,
where superscript $n$ indicates a last neutron in $^{10}$Li.
From Fig.~\ref{potential10}, it is found that the Pauli-blocking acts repulsively for the $p$-wave, 
which explains the inversion phenomenon in $^{10}$Li.

We briefly mention the experimental situation of $^{10}$Li.
The experimental identification of the $s$-wave and $p$-wave states in the $^9$Li+$n$ threshold energy region are difficult
and the conclusive understanding of the $^{10}$Li spectroscopy is still an open problem \cite{Je06}.
In the recent observation \cite{Je06}, $E_r\simeq0.38$ MeV with the decay width $\Gamma\simeq0.2$ MeV is suggested as the $p$-wave resonance, which is rather close to our results for the $1^+$ state.
For the $s$-wave state, $a_s=-13\sim-24$ fm is suggested, which agrees with our results of the $2^-$ state.
It is highly desired that further experimental data are obtained for the low-lying states of $^{10}$Li.

\subsection{$^{11}$Li}

We solve $^{11}$Li in a coupled $^9$Li+$n$+$n$ model
and show the detailed properties of $^{11}$Li in Table~\ref{tab:11}, 
the partial wave components $P((nlj)^2)$ for halo neutrons, the various r.m.s. radius, the relative distance between halo neutrons ($R_{n\mbox{-}n}$) and the core-$2n$ distance ($R_{c\mbox{-}2n}$), and the expectation value of the opening angle between two neutrons $\theta$
measured from the $^9$Li core, respectively.
The case using TOSM for $^9$Li gives a large $P((1s)^2)$ value, comparable to $P((p_{1/2})^2)$
for halo neutrons and a large matter radius $R_m$ for $^{11}$Li, which are enough to explain the observations.
The case of ``Inert Core'' gives small $P((1s)^2)$ and small $R_m$ values, which disagree with the experiments.
From the difference between two models, it is found that the tensor and pairing correlations in $^9$Li play important roles to break the magicity and make the halo structure of $^{11}$Li, in addition to the $s$-$p$ inversion phenomenon in $^{10}$Li.
As was shown in the previous study\cite{My07b}, the Pauli blocking effect from the tensor correlation is stronger than the pairing one to break the magicity of $^{11}$Li.

\begin{table}[t]
\caption{Ground state properties of $^{11}$Li with $S_{2n}=0.31$ MeV, where two kinds of the $^9$Li descriptions of TOSM and Inert Core are shown, respectively. 
Details are described in the text.}
\label{tab:11} 
\begin{center}
\begin{ruledtabular}
\renc{\baselinestretch}{1.15}
\begin{tabular}{c|cccc|c}
                           &   TOSM   & Inert Core & Expt. \\
\noalign{\hrule height 0.5pt} 				    
$P((p_{1/2})^2))$ [\%]     &  $42.7$  &  90.6      & ---  \\
$(1s_{1/2})^2 $            &  $46.9$  &   4.3      & 45$\pm$10\cite{Si99} \\
$(p_{3/2})^2  $            &  $2.5$   &  $0.8$     & ---  \\
$(d_{3/2})^2  $            &  $1.9$   &  $1.3$     & ---  \\
$(d_{5/2})^2  $            &  $4.1$   &  $2.1$     & ---  \\
$(f_{5/2})^2  $            &  $0.5$   &  $0.2$     & ---  \\
$(f_{7/2})^2  $            &  $0.6$   &  $0.3$     & ---  \\
\noalign{\hrule height 0.5pt}
$R_m $ [fm]                &  3.41    &   2.99     & 3.12$\pm$0.16\cite{Ta88b} \\
                           &          &            & 3.53$\pm$0.06\cite{To97} \\
                           &          &            & 3.71$\pm$0.20\cite{Do06} \\
$R_p $                     &  2.34    &   2.24     & 2.88$\pm$0.11\cite{Ta88b} \\
$R_n $                     &  3.73    &   3.23     & 3.21$\pm$0.17\cite{Ta88b} \\
$R_{ch}   $                &  2.44    &   2.34     & 2.467$\pm$0.037\cite{Sa06} \\
                           &          &            & 2.423$\pm$0.034\cite{Pu06} \\
$R_{n\mbox{-}n}  $         &  7.33    &   6.43     & \\
$R_{c\mbox{-}2n} $         &  5.69    &   4.26     & \\
\noalign{\hrule height 0.5pt}
$\theta   $ [deg.]         & 65.3     &  73.1      & \\
\end{tabular}
\end{ruledtabular}
\end{center}
\end{table}

\begin{figure}[t]
\centering
\includegraphics[width=8.0cm,clip]{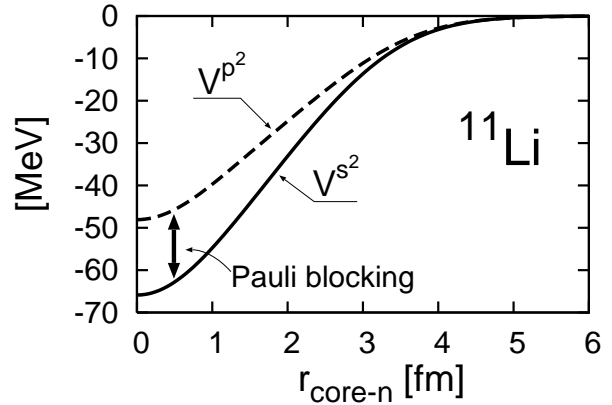}
\caption{Effective $^9$Li-$n$ potentials, $V^{s^2}$ and $V^{p^2}$ obtained by renormalizing the core excitation in $^{11}$Li,
where the $^9$Li-$n$ potentials are transformed into equivalent ones using the inert $^9$Li core.
$V^{s^2}$ is taken to be the same as the original folding potential $V_{cn}$.}
\label{potential11}
\end{figure}

Here, similar to the $^{10}$Li system, we show the Pauli-blocking effect in $^{11}$Li using the effective $^9$Li-$n$ potential 
and discuss the difference of the blockings between $^{10}$Li and $^{11}$Li.
In Fig.~\ref{potential11}, we depict the effective $^9$Li-$n$ potentials $V^{s^2}$ for the $(1s)^2$ configuration of $^{11}$Li
and $V^{p^2}$ for the $(0p)^2$ configuration, which are derived as follows;
We use the inert description of the $^9${Li} core, namely, no Pauli-blocking from the tensor and pairing correlations
and renormalize this blocking effect into $V^{s^2}$ and $V^{p^2}$.
For $s$-waves, $V^{s^2}$ is given by the original folding potential $V_{cn}$ in Eq.~(\ref{H10}),
which is also the same as $V^{s}$ in Fig.~\ref{potential10}.
For $p$-waves, we reduce the strength of $V_{cn}$ to simulate the blocking effects on the $(0p)^2$ configuration as follows;
First, we take only the $(0p)^2$ configuration for $\chi(nn)$ in Eq.~(\ref{OCM11}) using TOSM for $^9$Li, 
which gives $(E_r,\Gamma)=(1.50, 0.65)$ MeV of the $^{11}$Li ground state energy.
Second, we adopt the inert $^9$Li core and adjust $V_{cn}$ to reproduce the above resonance energy, which gives $V^{p^2}$.
The difference, $\Delta_{PB}^{n^2}$, between $V^{p^2}$ and $V^{s^2}$ reflects the strength of the Pauli-blocking for each last neutron in $^{11}$Li:
\begin{eqnarray}
  V^{p^2}(r) &\sim& V_{cn}(r) + \Delta_{PB}^{n^2}(r),
  \\
  V^{s^2}(r) &\sim& V_{cn}(r),
\\
\Delta_{PB}^{n^2}(r)&\sim&  V^{p^2}(r)-V^{s^2}(r),
\end{eqnarray}
where the superscript $n^2$ indicates the last two neutrons in $^{11}$Li.
In Fig.~\ref{potential11}, it is found that the Pauli-blocking acts repulsively on the $(0p)^2$ configuration of $^{11}$Li
and makes the $(0p)^2$ and $(1s)^2$ configurations placed at the degenerated energies, so that they can couple by the pairing interaction.
The energy difference between two configurations is $-0.1$ MeV in $^{11}$Li obtained in the previous study\cite{My07b}.
In comparison with the $^{10}$Li system shown in Fig.~\ref{potential10}, 
it is found that the blocking effect $\Delta_{PB}^{n^2}$ in $^{11}$Li is stronger than $\Delta_{PB}^{n}$ for $^{10}$Li.
This difference comes from the blocking of the tensor correlation in $^{11}$Li being stronger than the $^{10}$Li case, 
due to the presence of the last two neutrons in the $p_{1/2}$ orbit, as was explained in Sec.~\ref{sec:PB}.

\begin{table}[t]
\caption{Internal energy difference between $^9$Li inside the $^{11}$Li ground state
and the isolated $^9$Li ground state in units of MeV.}
\label{ene_diff}
\begin{center}
\begin{ruledtabular}
\renc{\baselinestretch}{1.15}
\begin{tabular}{cccccc}
$\Delta E$  &  Kinetic    &  Central  & Tensor  & LS       & Coulomb   \\
\noalign{\hrule height 0.5pt}
     2.0    &  $-0.8$     &  0.5      & 2.7     & $-0.3$   &  0.0004   \\
\end{tabular}
\end{ruledtabular}
\end{center}
\end{table}

In Table \ref{ene_diff}, we see the structure change of $^9$Li inside $^{11}$Li from the relative energy difference $\Delta E$
between the $^9$Li core inside $^{11}$Li and the isolated $^9$Li ground state.
The differences of each component in the $^9$Li Hamiltonian are also shown.
The positive value indicates the energy loss.
Among the Hamiltonian of $^9$Li, it is found that the energy loss of the tensor force is the largest. 
This is caused by the Pauli-blocking effect on the tensor correlation of $^9$Li in $^{11}$Li.
Corresponding to this, the kinetic energy is gained due to the reduction of the high momentum 
component produced by the tensor force.
Totally, the $^9$Li cluster is excited in energy in the $^{11}$Li ground state.

\begin{figure}[t]
\centering
\includegraphics[width=8.4cm,clip]{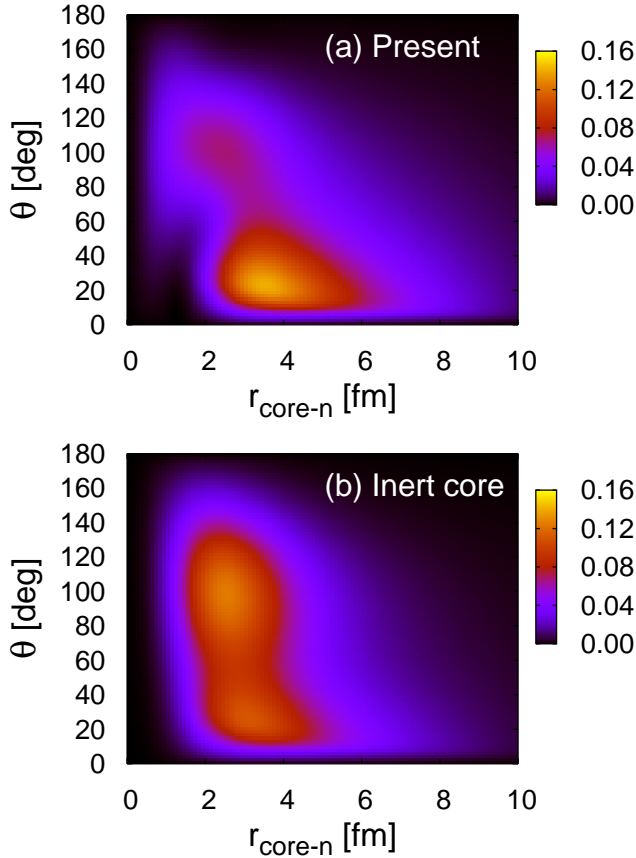}
\caption{(Color online) Two neutron correlation density $\rho_{nn}(r_{{\rm core}-n},\theta)$ for $^{11}$Li. 
(a) is the present calculation using TOSM of $^9$Li and (b) is the Inert Core case of $^9$Li, respectively.}
\label{fig:density}
\end{figure}

The spatial correlations of the halo neutrons in $^{11}$Li is interesting\cite{Es92,Zh93,Ni01,Ha05} 
and shown in Fig. \ref{fig:density}.
We depict the density distribution of halo neutrons $\rho_{nn}(r,\theta)$ in $^{11}$Li 
as functions of $^9$Li-$n$ distance, $r$, and the opening angle between two neutrons, $\theta$, with the following definition; 
\begin{eqnarray}
    \rho_{nn}(r,\theta)
&=& \int_0^\infty dr^\prime  \rho_{nn}(r,r^\prime,\theta)
    \\
    \rho_{nn}(r,r^\prime,\theta)
&=& 8\pi^2r^2 {r^\prime}^2 \sin\theta\
    \nonumber
    \\
&\times& \bra \Psi^J(^{11}{\rm Li},r,r',\theta) | \delta(r-r'')\delta(r'-r''')
    \nonumber
    \\
&\times& \delta(\theta-\theta')| \Psi^J(^{11}{\rm Li},r'',r''',\theta')\ket,
    \label{density}
\end{eqnarray}
where the total wave function of $^{11}$Li has three variables of the $^9$Li-$n$ distances with $r$ and $r^\prime$ for
each neutron and their opening angle $\theta$ and in Eq.~(\ref{density}), we integrate only the variables of the ket part. 
In the present TOSM case, (a), it is confirmed that the di-neutron type configuration ( a large $r$ and a small $\theta$) 
gives a maximum value of the density, although the density of neutrons is widely distributed.
Contrastingly, the Inert Core case, (b) with a small $s^2$ component does not show much enhancement of the di-neutron type configuration
and the cigar type configuration (a small $r$ and a large $\theta$) coexists with it, 
which is similar to the other halo nuclei, such as $^6$He\cite{Zh93}.
These two results indicate the role of the $s^2$ component on the formation of the di-neutron type configuration as follows: 
The $s^2$ component in the $^{11}$Li ground state increases the amplitude of the tail region of two neutrons far from $^9$Li,
and these neutrons tend to come close to each other to gain the interaction energy between them.
As a result, the di-neutron type configuration is enhanced, although the spatial distribution of neutrons are still wide.
The spatial distribution of two neutrons also affects the opening angle $\theta$, 
where the present TOSM case shows a smaller $\theta$ value than the Inert Core one, as shown in Table \ref{tab:11}.

The large $s^2$ component in $^{11}$Li enhances the core-$2n$ distance, which 
affects not only the matter radius but also the proton radius of $^{11}$Li,
because the proton radius of $^{11}$Li consists of the core part and the core-$2n$ distance ($R_{c\mbox{-}2n}$)
in the following form;
\begin{eqnarray}
R^2_p(^{11}{\rm Li})&=&R^2_p(^{9}{\rm Li})+\left(\frac{2}{11}\right)^2 R_{c\mbox{-}2n}^2.
\end{eqnarray}
The latter term represents the recoil effect of the c.m. motion from the halo neutrons.
Experimentally, the charge radius of $^{11}$Li was measured recently and its value is enhanced from the one of $^9$Li\cite{Sa06,Pu06}, as shown in Table \ref{tab:11}.  
The present wave functions provide reasonable value for $^{11}$Li.
This enhancement is mainly caused by the large distance between $^9$Li and the paired $2n$.
The Inert Core case gives the small charge radii for $^{11}$Li due to the small distance of $R_{c\mbox{-}2n}$, 
since the $s^2$ component is very small.

\begin{figure}[t]
\centering
\includegraphics[width=8.4cm,clip]{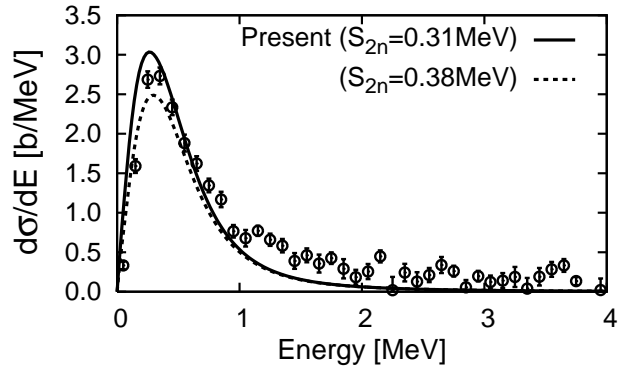}
\caption{Calculated Coulomb breakup cross section measured from the $^9$Li+$n$+$n$ threshold energy
in comparison to the experiments\cite{Na06}. There are two lines with different $S_{2n}$ for the $^{11}$Li ground state.}
\label{fig:cross}
\end{figure}

We show the three-body Coulomb breakup strength of $^{11}$Li from the ground state into the $^9$Li+$n$+$n$ system and compare it with the new data by the RIKEN group\cite{Na06} in Fig.~\ref{fig:cross}. 
We use the Green's function method combined with the complex scaling method to calculate the three-body breakup strength\cite{My03,My07b} using the dipole strength and the equivalent photon method, where experimental energy resolution is taken into account\cite{Na06}.
In Fig.~\ref{fig:cross}, it is found that the present model reproduces the experiment, in particular, for the low energy peak
observed at around 0.5 MeV.
For reference, we show the other results using a different two-neutron separation energy $S_{2n}$ of the $^{11}$Li ground state as 0.38 MeV suggested by the recent experiment\cite{Ba04}. Here we adjust the parameter $\delta$ in the $^9$Li-$n$ potential.
In that case, the matter radius of $^{11}$Li is 3.28 fm.
In the strength, only the magnitude at the peak is slightly reduced from the present results using $S_{2n}$=0.31 MeV
due to the threshold effect from the halo structure in the ground state\cite{Na94,My01}.
Seeing more closely, our results seem to underestimate the cross section at $E>1$ MeV, which is commonly seen in the results with different $S_{2n}$.

\begin{figure}[t]
\centering
\includegraphics[width=8.4cm,clip]{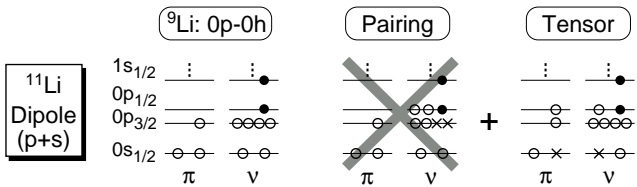}
\caption{Schematic illustration for the Pauli-blocking in the dipole states in $^{11}$Li
corresponding to the three kinds of the $^9$Li configurations.
Details are described in the text.}
\label{fig:Pauli2}
\end{figure}

Several three-body calculations of $^{11}$Li predicts the dipole resonances in the low excitation energy region with 
the decay widths, relatively smaller than the resonance energies\cite{Ga02b,Ku02}.
In fact, when we assume the inert $^9$Li core and take a phenomenological $^9$Li-$n$ potential with a deep $s$-wave potential, the dipole resonances of $1/2^+$, $3/2^+$ and $5/2^+$ are obtained\cite{My07b}.
Contrastingly, in the present model of $^{11}$Li with the tensor and pairing correlations, we cannot find any three-body dipole resonances\cite{My07b}.
Here, we discuss why the three-body dipole resonances are not obtained in the present model. 
To do this, we consider the Pauli-blocking effect on the dipole excited states in the analogy of the ground state as was explained in Sec.~\ref{sec:PB}.
In Fig. \ref{fig:Pauli2}, we show the dominant shell model type configuration for the dipole excited states, in which 
the last two neutrons occupy $1s_{1/2}$ and $0p_{1/2}$ orbits, respectively.
In this configuration, due to the Pauli-principle, the blocking effect for the tensor part becomes weaker than that in the $p^2$ configuration of the ground state, shown in Fig. \ref{fig:Pauli}(d).
This weakness of the blocking reduces the repulsive effect on the $p$-orbit and makes the energy gap between 
$0p_{1/2}$ and $1s_{1/2}$ orbits wider to get close to an usual order of the shell model orbits.
In this situation, the $(0p_{1/2})(1s_{1/2})$ configuration loses the energy, which means that the dipole resonances have relatively higher excitation energies than the phenomenological potential model.
Furthermore, since the $s$-wave neutron is easy to decay, the dipole resonance has a large decay width when the resonance energy is relatively high, which makes one difficult to find out its location in energy spectra to confirm the physical signature, such as the peak in the strength function\cite{My03}.

\begin{figure}[t]
\centering
\includegraphics[width=8.1cm,clip]{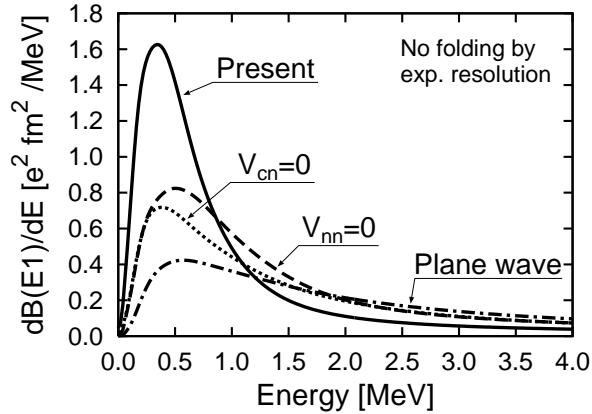}
\caption{Effect of the final state interactions on the $E1$ transition strength of $^{11}$Li.}
\label{fig:FSI}
\end{figure}

We further try to understand the physical origin of the low energy peak in the dipole strength around 0.5 MeV from the $^9$Li+$n$+$n$ threshold energy. 
Here, we investigate the effect of the final state correlations induced by the interactions between the $^9$Li+$n$+$n$ system.
Using TOSM for $^9$Li, in Fig.~\ref{fig:FSI}, we show the other three strength functions in addition to the original full calculation,
in which either the core-$n$ or $nn$ interactions is cut off, respectively, or both of them are cut off, namely, 
the plane wave treatment of the dipole excited states.
In these calculations, 
we do not change the ground state properties and do not take care of the experimental energy resolution of the strength.
The results show that the plane wave gives the small peak at 0.5 MeV in comparison with the present model, which indicates the importance of the final state correlations.
Among the final state correlations, the core-$n$ and $nn$ interactions give comparable contributions and 
both effects are necessary to reproduce the low energy peak in the TOSM case.
This fact means that both correlations of core-$n$ and $nn$ are important to explain 
the breakup mechanism of $^{11}$Li into the $^9$Li+$n$+$n$ system. 
The effects of the $p$-wave resonances and the virtual $s$-states of $^{10}$Li, and of the di-neutron type configuration
in the final states are interesting and would be investigated elsewhere.

\subsection{Electromagnetic properties of Li isotopes}

\begin{table}[t]

\caption{$Q$ moments of $^9$Li and $^{11}$Li in units of $e$ fm$^2$.}
\label{tab:Q}
\begin{ruledtabular}
\begin{tabular}{c|cc}
                           &  $^9$Li  & $^{11}$Li \\
\noalign{\hrule height 0.5pt} 
TOSM                       &  $-2.65$ &  $-2.80$  \\
AMD\cite{En04}             &  $-2.66$ &  $-2.94$  \\ 
SVM\cite{Va02}             &  $-2.74$ &  $-3.71$  \\
Experiment\cite{Ar94}      & $-2.74\pm 0.10$ & $ 3.12\pm 0.45$ ($|Q|$) \\
Experiment\cite{Bo05}      & $-3.06\pm 0.02$ &  --- \\
\end{tabular}
\end{ruledtabular}

\caption{$\mu$ moments of $^9$Li and $^{11}$Li in units of $\mu_N$.}
\label{tab:mu}
\begin{ruledtabular}
\begin{tabular}{c|cc}
                           &  $^9$Li  & $^{11}$Li \\
\noalign{\hrule height 0.5pt} 
TOSM                       &   3.69   &   3.77    \\
AMD\cite{En04}             &   3.42   &   3.76    \\
SVM\cite{Va02}             &   3.43   &   3.23    \\
Experiment\cite{Ar94}      &   3.44   &   3.67    \\
\end{tabular}
\end{ruledtabular}

\end{table}

We show the $Q$ and $\mu$ moments of the $^9$Li and $^{11}$Li in Tables \ref{tab:Q} and \ref{tab:mu},
where only the absolute value of the $Q$ moment of $^{11}$Li is reported in the experiments\cite{Ar94}.
The present model describes reasonably those values for $^9$Li and $^{11}$Li.
For $Q$ moments, the values of $^9$Li and $^{11}$Li do not differ so much to each other.
This result is similar to that of AMD\cite{En04} and different from SVM based on the multi-cluster model\cite{Va02}.
Here, similar to the charge radius, 
we discuss the recoil effect in the $Q$ moment of $^{11}$Li by expanding its operator $Q(^{11}{\rm Li})$
into the core part ($Q(^{9}{\rm Li})$), the last neutron part and their coupling part as
\begin{eqnarray}
    Q(^{11}{\rm Li})
&=& Q(^{9}{\rm Li})+ \sqrt{\frac{16\pi}{5}} 3 e \left(\frac{2}{11}\right)^2 {\cal Y}_{20}({\bf R}_{c\mbox{-}2n})
    \nonumber\\
&-& 8\pi \sqrt{\frac{2}{3}} [O_1(^9{\rm Li}),{\cal Y}_1({\bf R}_{c\mbox{-}2n})]_{20},
    \\
    O_{1m}(^9{\rm Li})
&=& e \sum_{i\in {\rm proton}} {\cal Y}_{1m}({\bf a}_i),
\end{eqnarray}
where ${\cal Y}_{lm}({\bf r})\equiv r^l Y_{lm}(\hat{\bf r})$ and $\{{\bf a}_i\}$ are the internal coordinates of protons in $^9$Li.
In our wave function of $^{11}$Li, last two neutrons almost form the $0^+$ state with the probability of around 99\%.
In that case, the $Q$ moment for the relative motion part of $^9$Li-$2n$($c$-$2n$) having the relative coordinate ${\bf R}_{c\mbox{-}2n}$ 
almost vanishes because of the non-zero rank properties of the $Q$ moment operator. 
This means that the recoil effect from the clusterization is negligible. 
Therefore, the $Q$ moment of $^{11}$Li is caused mainly by the $^9$Li core part inside $^{11}$Li.
In our wave function, the spatial properties of the proton part of $^9$Li inside $^{11}$Li do not change so much.
Hence, the $Q$ moment of $^{11}$Li is similar to the value of the isolated $^9$Li.
Small enhancement from $^9$Li to $^{11}$Li mainly comes from the lacking of the high momentum component of the tensor correlation 
due to the Pauli-blocking in $^{11}$Li, which extends the radial wave function of $^{11}$Li.
The experimental information of the $Q$ moment is important to understand the structure of $^{11}$Li.
It is highly desired that further experimental data are available for the $Q$ moment of $^{11}$Li.

For the $\mu$ moment, the value in $^{11}$Li is almost the Schmidt value of 3.79 $\mu_N$ of the $0p_{3/2}$ proton.
In $^9$Li, the $p_{1/2}$ proton is slightly mixed, which decrease the $\mu$ moment.
In $^9$Li, this $p_{1/2}$ proton is excited from the $0p_{3/2}$ orbit in a pair with the $p_{1/2}$ neutron 
and the $2p$-$2h$ excitation is Pauli-blocked in $^{11}$Li due to the additional neutrons.
As a result, the excitation of $p_{1/2}$ is suppressed, which makes the $\mu$ moment of $^{11}$Li close to 
the Schmidt value of the $p_{3/2}$ orbit.
The tendency of the increase of the $\mu$ moment from $^9$Li to $^{11}$Li can be obtained also in the shell model analysis using various effective interactions\cite{Su03}.

\section{Summary}\label{summary}

In summary, we have investigated the structures of $^{9,10,11}$Li systematically based on the extended three-body model,
in which the tensor and pairing correlations are explicitly considered for the $^9$Li part.
It is found that the two correlations in $^9$Li play an essential role to explain the anomalous structures of $^{10,11}$Li. 
The obtained results clearly demonstrate that the inert core assumption of $^9$Li is not reliable.
For $^{10,11}$Li, the Pauli-blocking induced by the tensor and pairing correlations in $^9$Li naturally 
explains not only the inversion phenomena of $^{10}$Li, but also the breaking of magicity and the halo formation in $^{11}$Li
consistently.
It is pointed out that the strengths of the Pauli-blocking between $^{10}$Li and $^{11}$Li including their excited states are different.
The interplay between the tensor correlation of $^9$Li and the halo neutrons mainly gives rise to this difference
and due to this difference, the blocking acts strongly for the $^{11}$Li ground state.
We have further investigated the various physical quantities of Li isotopes.
In $^{11}$Li, it is shown that the development of the halo structure induces the di-neutron type configuration, 
and the large matter and charge radius. 
The electromagnetic properties of $^9$Li and $^{11}$Li are reproduced in the present model.
The present model explains further the recent observation of the Coulomb breakup strength of $^{11}$Li. 
The low-energy peak in the Coulomb breakup strength originates from the final state interactions.

\begin{acknowledgments}
The authors thank Prof. H. Horiuchi for continuous encouragement and Prof. T. Nakamura for valuable discussions on experiments. 
This work was performed as a part of the ``Research Project for Study of
Unstable Nuclei from Nuclear Cluster Aspects (SUNNCA)'' at RIKEN and
supported by a Grant-in-Aid from the Japan Society for the Promotion of Science (JSPS, No. 18-8665)
and also by the JSPS Core-to-Core Program.
Numerical calculations were performed on the computer system at the Research Center for Nuclear Physics, Osaka University.
\end{acknowledgments}

\end{document}